\documentclass[russian,aps]{revtex4}
\usepackage{babel}
\usepackage[cp1251]{inputenc}
\usepackage{graphicx}
\usepackage{amsfonts,amssymb,eucal,amsmath}

\begin{document}

\title{Realization of the anomalous scattering method in crystallography on the basis of
the parametric X-ray radiation}

\author{I.D.Feranchuk and A.P.Ulyanenkov}

\address{Byelorussian State University, F.Skariny Av., 4, 220050 Minsk,
Republic of Belarus}

\begin{abstract}
Spectra of parametric X-ray radiation (PXR)  are considered in the range of anomalous dispersion for one of
the atoms in a crystal elementary cell. The PXR spectra are calculated both for the ultra-relativistic ($E
\geq 50 MeV$) and non-relativistic  ($E \simeq 100 KeV$) electrons for the specific conditions corresponding
to the organic compounds . It is shown that analysis of the  RXR angular distribution permits one to realize
the method of anomalous scattering for direct measurement of the phases of structure amplitudes without
additional isomorphic intrusion of a heavy atom.
\end{abstract}

\maketitle


\section{Introduction}

Development of the physical methods for direct measurement of the phases of structure amplitudes in the
X-ray analysis of the crystals has always attracted a great attention \cite{Woolfson}. Solution of the phase
problem is especially important for determination of the structure of the complicated organic compounds with
elementary cell including a lot of atoms. Quite long ago the effective and simple methods of the direct
measurement of phases  in the X-ray structure analysis were suggested and realized for some special cases
(for example, \cite{Vainshtein}). The first one is the method of double isomorphic replacement when the
heavy atom strongly scattered X-ray radiation is introduced to the investigated structure   \cite{Harker}.
Another approach is the anomalous scattering method (ASM) when the intensities of the diffraction reflexes
are measured  using the radiation with two wave lengths near the absorption edge of one the atoms in the
elementary cell of the analyzed crystal \cite{Bijvoet}. These methods have big advantage for possible
applications in the practical X-ray structure analysis because they permit one to measure the phases in
rather thin crystals when the kinematic diffraction theory is available and the observed data can be simply
interpreted  \cite{Porai} - \cite{Attfield} .

A series of methods for solution of the phase problem were suggested on the basis of the interference
phenomena in various variants of the many-beam diffraction. As a rule, the  wave interference is essential
when the dynamical effects are clearly appeared in the X-ray diffraction and the photons can scatter
multiply  in the analyzed crystal which should be perfect. Realization of such methods for mosaic and
macromolecular crystals is rather complicated problem which is developed actively by many authors at present
(for example, \cite{Hart} - \cite{Stetsko} and references therein).

However, the phase information in  many-beam diffraction experiments can be extracted from the diffraction
data by means of the complicated and precise additional measurements even in the ideal crystal. Therefore
elaboration of the effective and universal approach to the experimental realization of ASM is very actual
task because in this case only the intensities of reflexes should be measured and they are strictly
connected with the structure amplitudes and their phases in the framework of kinematical diffraction. There
are two main problems in this way: i) the reflex intensities should be measured with X-rays of various wave
lengths and ii) the absorption edge of  atoms in the most organic compounds corresponds to very soft X-ray
radiation (for example, $K_{\alpha}$-line for the sulphur atom  S, which is the heaviest atom in the most of
aminoacids, corresponds the energy of photons   $\hbar \omega_K \simeq 2.40 keV $, or wave length $
\lambda_K \simeq 5.13 A$). It makes difficult to use ASM for the X-ray structure analysis in home
laboratories because of lack of the soft X-ray sources with smoothly variable wave length. Possibly, these
problems could be overcome by means of synchrotron radiation but even in this case some difficulties are
appeared when choosing the crystal-monochromator for soft X-rays  \cite{Hart1}.

In the present paper we discuss another possibility to use ASM for the structure analysis of organic
compounds by means of measurement of  angular distribution of PXR generated by electrons passing through the
investigated crystal \cite{patent}. There are two variants in order to realize such approach in laboratory
conditions. One of them can be built on the basis of PXR from ultrarelativistic electrons  (with the energy
$E \sim 30 - 50 MeV$), when the investigated crystal is arranged inside the betatron. This type of PXR was
successfully obtained recently by authors of \cite{Piestrup}. In Sec.2 we will discuss the realization of
ASM in scope of such approach.

The other method (Sec.3) is based on PXR from the nonrelativistic electrons (with the energy $E \sim 100
keV$). The main characteristics and some applications of such form of PXR were considered in our papers
\cite{Fer1}-\cite{Fer4}.

\section{ASM on the basis of PXR from ultrarelativistic electrons}

Remind some qualitative features of the interaction of the ultrarelativistic electrons with crystals which
permit one to consider the PXR kinematics and most essential for preliminary analysis of the phenomenon
\cite{Bar83},\cite{Fer85}. Let us suppose that an electron with energy $E \gg m c^2$ (m is the electron
mass; c is the light velocity) enters the crystal with an arbitrary orientation relatively to the electron
velocity $\vec v$ (Fig.1). It is well known that the charge particle's electromagnetic field can be
represented as the pseudophoton beam   \cite{Bar83}, whose properties are closed to those of real photons
with the wave vectors  $\vec k = \omega \vec v/c^2$, concentrated in the angular cone  $\Delta \theta \simeq
mc^2/E$ along the vector $\vec v$ and the spectral density $n(\omega)$ that is defined by the following
formula \cite{Bar83}:

\begin{eqnarray}
\label{1} n(\omega)d\omega \simeq \frac{\alpha}{\pi \omega}\ln (\eta\frac{E}{\hbar \omega})d\omega,
\end{eqnarray}

\noindent where $\alpha = \frac{e^2}{\hbar c}\simeq \frac{1}{137}$ is the fine structure constant and $\eta$
is the the value of the order of unity.

\begin{figure} [ht]
\caption{ Sketch of the PXR experiment}
\includegraphics[width = 8cm,height =7cm]{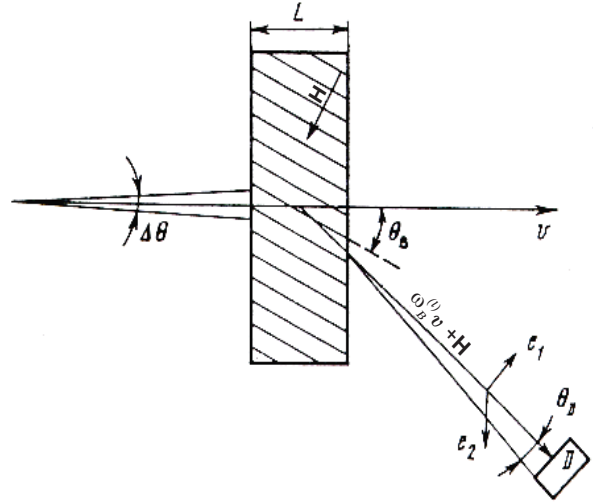}
\end{figure}

From this point of view the kinematics of the electromagnetic interaction between the electron and crystal
is equivalent to the interaction of the photon beam with the angle divergence  $\Delta \theta $ and the
"white" spectrum (\ref{1}) with the same crystal. In particular, the photons with the wave vector satisfied
to the Bragg condition with one of vector $\vec H$ of the reciprocal lattice of the crystal

\begin{eqnarray}
\label{2} (\vec k + \vec H)^2 \simeq k^2; \quad 2 \omega ( \vec v \vec H) + H^2 c^2 \simeq 0,
\end{eqnarray}

will be reflected by the corresponding crystallographic planes. It will lead to the appearance of the X-ray
radiation peak (reflex) over the angle  $2\theta_B$ with respect to the electron velocity. The value
$\theta_B$ is defined by the angle between the reflected planes and the vector  $\vec v$ (Fig.1), that is

$$
( \vec v \vec H) = - v  H\sin \theta_B.
$$

The totality of such reflexes conditioned by various crystallographic planes and distributed in the space
represents the integral spectrum of PXR from the ultrarelativistic electrons   \cite{Fer85}. Intensity of
the radiation in the every reflex is defined by the same structure amplitude $F(\vec H)$ as for scattering
of the real photons in the same direction. Besides, each reflex has the "fine structure" that is
characteristic angular and frequency distributions  \cite{Fer85} correspondingly near the angle $2\theta_B$
and frequency $\omega_B$ (wave length $\lambda_B$), which value can be found strictly from the formula
(\ref{2})

\begin{eqnarray}
\label{3}  \omega_B = \frac{H c^2}{2 v \sin \theta_B}; \quad \lambda_B = \frac{4\pi v \sin \theta_B}{H c}.
\end{eqnarray}

These properties of the PXR spectrum were confirmed by a lot of experiments (for example, \cite{Exper} and
references therein) and their results in a good agreement with the theory.

In accordance with the formula (\ref{3}) the radiation frequency in the PXR reflex can be varied by simple
rotation of the crystal relatively to the electron beam and it is the basis for development of new sources
of the monochromatic X-ray radiation with smoothly changeable frequency  \cite{Piestrup}, \cite{Fer1}. But
the main purpose of the present paper is the justification of the possibility to use this property of PXR
for solution of  another problem which is the direct measurement of the phase of structure amplitudes
$F(\vec H)$ on the basis of ASM.  It means that the incident angle of the electron in the crystal should be
chosen such a way that the radiation frequency in the PXR reflex from the analyzed plane was near the
frequency of absorption K-edge  $\omega_K$ of one of the substances forming the crystal

\begin{eqnarray}
\label{4} \sin \theta_K^H =   \frac{H c^2}{2 v \omega_K} = \frac{c}{4 \pi v} H \lambda_K,
\end{eqnarray}

\noindent where $\lambda_K$ is the wave length of $K_{\alpha}$-line for the considered substance  .

Analogous approach was also discussed before in our paper \cite{Fer89}. But this analysis was referred to
the case of X-rays corresponding to the characteristic frequencies of the heavy atoms and, besides, was
considered for the conditions of the dynamical diffraction in the perfect crystals. But both conditions are
not realized in the organic compound crystals considered in the present paper.

It is important to notice that the characteristic parameters which define the interaction of the X-rays with
the organic crystals are essentially differed from the analogous values for inorganic crystals. In
particular the former ones have a comparatively small susceptibility  $\chi_0$ and a quite large extinction
length $L_e$ which is the same order as the investigated sample size  L. In its turn, the value L is usually
smaller than the absorption length $L_{abs}$. In order to estimate these parameters and consider the
quantitative analysis of the effect let us choose for definiteness one of the typical organic crystals which
structure was recently investigated in the paper   \cite{krystal}. This is the crystal of the dimetil-ether
of the 4-fenilsemicarboacetil acid (DFA) with the structure formula

$$
C_{13} H_{17} N_3 O_4 S
$$

It has the monoclinic  singony and space group $P2_1/n$. The elementary cell parameters (a,b,c), its volume
$\Omega$ and the directing angle $\beta$ have the following values accordingly

\begin{eqnarray}
\label{5} a = 8.066, b = 15.812, c = 24.977 A; \quad \Omega = 3174 A^3; \quad \beta = 94.88.
\end{eqnarray}

Accordingly to the formula (\ref{3}) for the crystals with such values of the lattice parameters the wave
length of the most part of the PXR reflexes belongs to the range of  $\lambda \leq 20 A$. The only atom of
the sulphur S has the absorption edge in this range and we will consider it as the main object for
realization of ASM. One more reason for choice just S for this purpose is its universal spreading almost in
all organic crystals and its coordinate is found most precisely on the first stadium of the structure
analysis because of rather big charge \cite{Porai}.

Let us calculate the susceptibility $\chi_0$ \cite{James} of the considered crystal with the total number of
electrons in the elementary cell $Z_{tot}= 164$ and for the wave length  $\lambda_K \simeq 5.13 A$,
corresponding the K-edge of the atoms S

\begin{eqnarray}
\label{6} \chi_0 = - \frac{e^2 Z_{tot}\lambda^2_K }{4 \pi^2 m c^2 \Omega} \quad a = 8.066, b = 15.812, c =
24.977 A; \quad \Omega = 3174 A^3; \quad \beta = 94.88.
\end{eqnarray}

Then the unknown values are

\begin{eqnarray}
\label{7} \chi_0 \simeq - 1.23  10^{-5}; \quad L_e =  \frac{\lambda_K }{\pi|\chi_0 | } \simeq 13.3  \mu k;
L_{abs} \simeq 1 mm.
\end{eqnarray}

These values can be compared with the analogous ones for the crystal Si which has the susceptibility $\chi_0
\simeq 1.5 10^{-4}$ for the same wave length and essentially less extinction length.

Besides, the average effective charge $Z_{eff}$ for one atom is rather small for the organic crystals (for
the DFA crystal $Z_{eff} = 4$), and multiple scattering in these materials is essentially less than in the
inorganic crystals.

\section{Calculation of the phases of the structure amplitudes}

It can be shown that for the above mentioned parameters of the organic crystals  intensity of the PXR reflex
can be calculated in the scope of the kinematical approximation when it is defined by means of the following
simple formula

\begin{eqnarray}
\label{8} I_{\vec H,\omega} = A(E) |F(\vec H,\omega )|^2 ,
\end{eqnarray}

\noindent where the value $A(E)$ is the universal coefficient which  depends on the particle energy but
doesn't connect with  the crystal structure \cite{Fer85}.

So, one can measure the value $|F(\vec H,\omega )|$ strictly from the intensity measurements

\begin{eqnarray}
\label{9} |F(\vec H,\omega )| = \sqrt{\frac{I_{\vec H,\omega}}{ A(E) }}.
\end{eqnarray}

From the other side, in the of the anomalous dispersion this value can be expressed as follows

\begin{eqnarray}
\label{10} |F(\vec H,\omega )|^2 = | F_0(\vec H ) + [\Delta f' (\omega) + i \Delta f'' (\omega)] S(\vec
H )|^2,
\end{eqnarray}

where $F_0(\vec H,\omega )$ is the value of the crystal structure amplitude calculated with the wave length
far from the range of the anomalous dispersion; $\Delta f' (\omega), \Delta f'' (\omega)$ are the anomalous
corrections to the atom amplitude which are essential in the range $\omega \simeq \omega_K$ and their values
are well known \cite{Vainshtein} ; $S(\vec H,\omega )$ is the structure factor of the sulphur atoms in the
elementary cell.

The relative interval of the anomalous dispersion $\Delta \omega_K/\omega_K$ is the value of the order of
unit  \cite{Vainshtein} and it is essentially wider than the characteristic frequency interval $\Delta
\omega_B/\omega_B \simeq m c^2/E \ll 1$ for the PXR reflex. So, one can measure the values $F_{1,2}(\vec
H,\omega_{1,2} )$ for two different frequencies in the interval of the anomalous dispersion. From the other
side, one can also use the reflex corresponding to the reciprocal lattice vector $(\bar \vec H = - \vec H)$,
as it is usually considered in the ASM. This reflex can measured in the result of the rotation of the
crystal on the $180^o$ around the axis perpendicular to the electron velocity and with the same value A(E).
In this case the structure amplitude is defined by the equation

\begin{eqnarray}
\label{11} |F(\bar \vec H,\omega )|^2 = | F^*_0(\vec H) + [\Delta f' (\omega) + i \Delta f''
(\omega)] S^*(\vec
H)|^2.
\end{eqnarray}

This value can also be measured for the same frequencies $\omega_{1,2} )$. In the result 4 equations
will be obtained for 2 unknown values $F_0(\vec H ),S(\vec H )$ and their phases which can be uniquely
calculated.

\section{Conclusions}

So, the method of the direct measurement of the phases of the structure amplitudes on the basis of the
parametric X-ray radiation in the range of the anomalous dispersion is justified theoretically.

\section{Acknowledgments}

Authors  are grateful to the International Scientific Technical Center (Grants B-626) for the support of
this work.


\end{document}